\newcommand{\ket}[1]{\left| #1 \right\rangle}

\newcommand{\qo}[1]{``#1''}
\documentclass[12pt]{iopart}

\usepackage[toc,page]{appendix}
\usepackage{graphicx,bm,color,mathptmx,hyperref}
\usepackage{amsfonts}
\begin{document}
\title[Quantum probabilities from quantum entanglement]{Quantum probabilities from quantum entanglement: Experimentally unpacking the Born rule}
\author{J\'er\'emie Harris$^{1,2}$}
\address{$^1$Department of Physics, University of Ottawa, 25 Templeton, Ottawa, Ontario, K1N 6N5 Canada}
\address{$^2$The Max Planck Centre for Extreme and Quantum Photonics, University of Ottawa, Ottawa, Ontario, K1N 6N5, Canada}

\author{Fr\'ed\'eric Bouchard$^{1,2}$}
\address{$^1$Department of Physics, University of Ottawa, 25 Templeton, Ottawa, Ontario, K1N 6N5 Canada}
\address{$^2$The Max Planck Centre for Extreme and Quantum Photonics, University of Ottawa, Ottawa, Ontario, K1N 6N5, Canada}

\author{Enrico Santamato$^{3}$}
\address{$^3$Dipartimento di Fisica, Universit\`a di Napoli Federico II, Compl. Univ. di Monte S. Angelo, via Cintia, 80126 Napoli, Italy}

\author{Wojciech H. Zurek$^{4}$}
\address{$^4$Theoretical Division, Los Alamos National Laboratory, Los Alamos, NM 87545, USA}

\author{Robert W. Boyd$^{1,2,5}$}
\address{$^1$Department of Physics, University of Ottawa, 25 Templeton, Ottawa, Ontario, K1N 6N5 Canada}
\address{$^2$The Max Planck Centre for Extreme and Quantum Photonics, University of Ottawa, Ottawa, Ontario, K1N 6N5, Canada}
\address{$^5$Institute of Optics, University of Rochester, Rochester, New York, 14627, USA}

\author[cor1]{Ebrahim Karimi$^{1,2}$}
\address{$^1$Department of Physics, University of Ottawa, 25 Templeton, Ottawa, Ontario, K1N 6N5 Canada}
\address{$^2$The Max Planck Centre for Extreme and Quantum Photonics, University of Ottawa, Ottawa, Ontario, K1N 6N5, Canada}
\ead{ekarimi@uottawa.ca}

%
\vspace{10pt}
\begin{indented}
\item[]November 2015
\vspace{20pt}
\end{indented}

\begin{abstract}
The Born rule, a foundational axiom used to deduce probabilities of events from wavefunctions, is indispensable in the everyday practice of quantum physics. It is also key in the quest to reconcile the ostensibly inconsistent laws of the quantum and classical realms, as it confers physical significance to reduced density matrices, the essential tools of decoherence theory. Following Bohr's Copenhagen interpretation, textbooks postulate the Born rule outright. However, recent attempts to derive it from other quantum principles have been successful, holding promise for simplifying and clarifying the quantum foundational bedrock. A major family of derivations is based on envariance, a recently discovered symmetry of entangled quantum states. Here, we identify and experimentally test three premises central to these envariance-based derivations, thus demonstrating, in the microworld, the symmetries from which the Born rule is derived. Further, we demonstrate envariance in a purely local quantum system, showing its independence from relativistic causality.
\end{abstract}

%
%
%
%
%
\section{Introduction}

For almost a century, we have understood the Universe to obey the laws of quantum mechanics at microscopic scales~\cite{planck:14,heisenberg:27}. Quantum mechanics is arguably the most successful physical theory ever developed. Its validity in the microscopic regime has been affirmed by decades of pointed and rigorous experimental testing. Almost paradoxically, however, the laws of quantum mechanics, na\"ively interpreted, seem to prescribe behaviour incompatible with our day-to-day classical experience~\cite{schlosshauer:07,weinberg:13}.

Numerous quantum mechanical phenomena appear to defy classical explanation, including, for example, the existence of quantum state superpositions. Yet, the ostensible classicality of our familiar macroscopic world is undeniable: flagrant superpositions of classical states, such as that described by Erwin Schr\"odinger in his well-known cat-in-a-box thought experiment~\cite{schrodinger:35}, are simply never observed. Rather, our experience is one of absolutes: we observe a light bulb to be on or off, but certainly not on and off simultaneously, for example. 

The Born rule establishes a connection between the wavefunction used to represent the state of a quantum system (a purely mathematical object), and the probabilistic outcomes of measurements made on that system~\cite{born:26}, as experienced by observers (the \qo{physical reality}). {For a pure (fully coherent) physical system described by a wavefunction $\ket{\psi} = \sum_n c_n \ket{n}$, where $\ket{n}$ is a ``classically observable'' state the Born rule states that the probability} of witnessing a measurement outcome $|n\rangle$ is given by $|c_n|^2$. Moreover, the Born rule justifies averaging over (\qo{tracing out}) the environment, thereby validating the statistical interpretation of the reduced density matrices used to study decoherence. Therefore, it plays a central role in the theory of decoherence, which accounts for the emergence of classical behaviour from quantum substrates~\cite{schlosshauer:07}.  While one could adopt the Born rule as an axiom, following the Copenhagen interpretation and most textbooks, a physically transparent derivation would contribute greatly to clarifying the foundations of quantum mechanics. Given its importance to quantum mechanics, and to decoherence theory in particular, a satisfactory explanation of the quantum-to-classical transition depends upon one's ability to derive the Born probability rule (BPR) from simpler quantum mechanical principles. 

Attempts to reason up to the Born rule in this way have historically frustrated physicists and philosophers in equal measure~\cite{footnote:00}; indeed, the absence of universally satisfactory derivations of the BPR has led more pragmatic workers to resign themselves to outright postulation of the Born rule~\cite{footnote:01}. Such a position is tantamount to abandoning any attempt to explain, on fundamental grounds, how quantum theory might possibly correspond to physical reality, and, as such, is an unpalatable option to many. 

Relatively recently, the BPR was shown to arise as a consequence of certain properties, which are known to arise from the entanglement of quantum states~\cite{zurek:03a}. The central property, a symmetry known as \textit{envariance}, has been subjected to tests in its nonlocal form~\cite{vermeyden:15}, but until now, there has been no experimental substantiation of the other premises underlying this theoretical argument for the BPR. In other words, we know that entangled quantum states have symmetries that imply the Born rule, but we do not know whether physical systems indeed respect these (often counterintuitive) symmetries. The situation is similar to the \qo{EPR paradox} and Bell's theorem: its violation was predicted by quantum theory, but experimental tests were needed to ascertain that microsystems in our Universe really behave like this!

Here, we demonstrate experimentally the validity of three key premises required by this envariance-based BPR derivation, and thereby show that the most controversial logical \qo{ingredients} of the proof are, in fact, physically sound. {In particular, we experimentally verify 1) that pure quantum states consisting of two maximally entangled degrees of freedom are left unaltered by the action of successive ``swapping'' operations, each of which are carried out on a different (entangled) degree of freedom; 2) that these ``swapping'' operations cannot be noticed by observers having access to only one degree of freedom possessed by the overall quantum state; and 3) that kets defined over different Hilbert spaces that are linked to one another via tensor product are detected together upon measurement. We follow a variant of the original envariance argument to show that the three propositions above lead to the conclusion that the Born rule must hold in general. Further, since we experimentally test these propositions both for fully local and fully nonlocal quantum states, we demonstrate the key ingredients of the envariance argument in full generality, and allow the consequences of envariance to be distinguished from those of nonlocality for the first time.}

We note also that, prior to this study, the Born rule could only be tested for specific cases, rather than in its general form: a test of the Born rule would necessarily entail the repeated generation and measurement of a particular quantum state, and the comparison of the corresponding measurement  statistics to the predictions that one would obtain from the Born rule. The approach presented here provides the first means by which the \emph{general} validity of the Born rule can be verified from carrying out a series of well-defined experiments on a particular quantum state. These experiments are designed so as to test premises from which the Born Rule can be \emph{derived}.

\section{Theory}
{Any process of quantum measurement necessarily involves an interaction between two or more systems. That this is true can be seen by considering that a measurement, by definition, involves a transfer of information \textit{between} systems: system ${\cal S}$ has been \qo{measured} only if there exists another system ${\cal E}$, that carries information about the state of ${\cal S}$. For this reason, any theory of quantum measurement must necessarily describe an interaction between multiple degrees of freedom, which can be said to \qo{measure one another}. With these preliminaries in place, we now begin by briefly} reviewing the derivation of the Born rule, which motivates this investigation, bearing in mind that the simple scenario discussed here is readily generalized to the case of arbitrary quantum states (see: Appendix A). In this approach, one considers a bipartite quantum state $|\psi_{\mathcal SE}\rangle$, formed from an interaction between a system $\mathcal{S}$ and its environment $\mathcal{E}$~{\cite{footnotedef}}, whose Hilbert spaces are respectively spanned by the orthonormal kets $\{|s_1\rangle$, $|s_2\rangle\}$, and $\{|\varepsilon_1\rangle$, $|\varepsilon_2\rangle\}$. This process, often referred to as a \textit{pre-measurement}, takes the form
%
\begin{eqnarray}\label{eq:PreMeasurement}
	\frac{1}{\sqrt{2}}\left(|s_1\rangle + |s_2\rangle\right) |\varepsilon_1\rangle \rightarrow \frac{1}{\sqrt{2}}\left(|s_1\rangle|\varepsilon_1\rangle + |s_2\rangle|\varepsilon_2 \rangle \right) = |\psi_{\cal SE}\rangle.
\end{eqnarray}
We note that the final state (\ref{eq:PreMeasurement}) is expressed in Schmidt form, so that each of its constituent system and environment kets are orthonormal. Under these conditions, the Born rule predicts that an experiment carried out on the state $|\psi_{\cal SE}\rangle$ will yield a result corresponding to the system state $|s_1\rangle$ with probability $|1/\sqrt{2}|^2=1/2$, and a result corresponding to system state $|s_2\rangle$ with equal probability. But how can this be justified from more basic quantum mechanical principles? 

One can imagine applying a unitary operator, $\hat{U}_{\cal{S}}=\hat{u}_{\cal{S}} \otimes \hat{\mathbb{I}}_{\cal{E}}$, which acts only on the system, and a second unitary operator $\hat{U}_{\cal{E}}=\hat{\mathbb{I}}_{\cal{S}} \otimes \hat{u}_{\cal{E}}$, which acts only on the environment, to the state $|\psi_{\cal SE}\rangle$. Here, we take $\hat{\mathbb{I}}_{\cal{S}}$ and $\hat{\mathbb{I}}_{\cal{E}}$ to represent identity operators over the system and environment spaces, and $\hat{u}_{\cal{S}}$ and $\hat{u}_{\cal{E}}$ to represent nontrivial operations respectively carried out over the system and environment spaces. It is argued that, if there exist operators $\hat{U}_{\cal{S}}$ and $\hat{U}_{\cal{E}}$, such that $\hat{U}_{\cal{E}} \hat{U}_{\cal{S}}|\psi_{\cal SE}\rangle=\hat{u}_{\cal{S}} \otimes \hat{u}_{\cal{E}}|\psi_{\cal SE}\rangle=|\psi_{\cal SE}\rangle$, any properties modified by $\hat{U}_{\cal{S}}$ ($\hat{U}_{\cal{E}}$) cannot be ascribed to the system (environment) alone, since changes in these properties can be reversed by the application of a unitary operator acting on a completely separate part of the wavefunction (in quantum mechanical parlance, on a completely separate Hilbert space). This implies that an observer having access to the system (environment) alone would be unable to detect any change in the state of the system either the environment resulting from the application of $\hat{U}_{\cal{S}}$ or $\hat{U}_{\cal{E}}$ to the global state $|\psi_{\cal SE}\rangle$. Any property affected by either of these operators is defined as being \textit{envariant} under $\hat{U}_{\cal{S}}$ and $\hat{U}_{\cal{E}}$. 

The argument proceeds by introducing the unitary \qo{swapping'} operators $\hat{u}_{\cal{S}}=|s_1\rangle\langle s_2| + |s_2 \rangle \langle s_1|$ and $\hat{u}_{\cal{E}}=|\varepsilon_1\rangle\langle \varepsilon_2| + |\varepsilon_2 \rangle \langle \varepsilon_1|$, which respectively cycle the state labels 1 $\longleftrightarrow$ 2 of the system and environment. These operators satisfy the envariance relation above, since $\hat{u}_{\cal S}\otimes\hat{u}_{\cal E} |\psi_{\cal SE}\rangle= |\psi_{\cal SE}\rangle$. As a result, the properties affected by $\hat{u}_{\cal S}$ and $\hat{u}_{\cal E}$ must be envariant, and cannot be attributed to either the system or the environment alone. 

To make the case for the BPR, we adopt the following notational convention: for an observer having access only to the Hilbert space to which belongs a state $|x\rangle$ (i.e., the system or environment spaces), the probability that state $|x\rangle$ is observed upon measurement is given by ${\cal P}\left(\ket{x}{\Big |}\ket{\psi}\right)$, where $|\psi\rangle$ is the full state of the entangled system-environment. The Born rule can then be derived as follows~\cite{zurek:05,schlosshauer:05b}:

\begin{enumerate}
\item The swapped and counterswapped state $\hat{U}_{\cal{E}} \hat{U}_{\cal{S}}|\psi_{\cal SE}\rangle$ is mathematically identical to the initial state $|\psi_{\cal SE}\rangle$. The states must therefore be physically equivalent as well, meaning that \textit{the statistics witnessed by observers having access to the combined system and environment of the states $|\psi_{\cal SE}\rangle$ and $\hat{U}_{\cal{E}} \hat{U}_{\cal{S}}|\psi_{\cal SE}\rangle$ must be indistinguishable} (Premise I). Hence, ${\cal P}\left(|s_1\rangle {\Big |} |\psi_{\cal SE} \rangle \right) = {\cal P}\left(|s_1\rangle {\Big |} \, \hat{U}_{\cal{E}} \hat{U}_{\cal{S}} |\psi_{\cal SE} \rangle \right)$.

\item Since the swapping operator $\hat{U}_{\cal{E}}$ acts only on the environment, it cannot affect the state of the system. Consequently, \textit{the application of the swapping operator $\hat{U}_{\cal{E}}$ cannot affect the statistics witnessed by a ``system-only'' observer} (Premise II). Hence, ${\cal P}\left(|s_1\rangle {\Big |} \hat{U}_{\cal{E}} \hat{U}_{\cal{S}} |\psi_{\cal SE} \rangle \right)={\cal P}\left(|s_1\rangle {\Big |}\hat{U}_{\cal{S}} |\psi_{\cal SE} \rangle \right)$.

\item The system-swapped state, $\hat{U}_{\cal{S}} |\psi_{\cal SE}\rangle=1/\sqrt{2}\left(|s_2\rangle |\varepsilon_1\rangle + |s_1\rangle|\varepsilon_2\rangle \right)$, directly links the states $|s_1\rangle$ and $|\varepsilon_2\rangle$. \textit{Kets appearing together in the Schmidt decomposition of the global state must also appear together upon measurement, and are hence \qo{perfectly correlated}}~\cite{footnote:02} (Premise III). Therefore, ${\cal P}\left(|s_1\rangle {\Big |}\hat{U}_{\cal S} |\psi_{\cal SE}\rangle \right) = {\cal P}\left(|\varepsilon_2\rangle {\Big |} \hat{U}_{\cal S} |\psi_{\cal SE}\rangle \right)$.

\item By Premise II, a swapping operator acting on the system cannot affect the statistics of the environment, and so ${\cal P}\left(|\varepsilon_2\rangle {\Big |} \hat{U}_{\cal S} |\psi_{\cal SE}\rangle \right)= {\cal P}\left(|\varepsilon_2\rangle {\Big |} |\psi_{\cal SE}\rangle \right)$.

\item Since $|\varepsilon_2\rangle$ is paired with $|s_2\rangle$ in $|\psi_{\cal SE}\rangle$, it follows from Premise III that ${\cal P}\left(|\varepsilon_2\rangle {\Big |} |\psi_{\cal SE} \rangle \right)={\cal P}\left(|s_2\rangle {\Big |} |\psi_{\cal SE} \rangle \right)$.
\end{enumerate}

Provided that the premises called upon by arguments 1-5 hold true, one concludes that ${\cal P}\left(|s_1\rangle {\Big |} |\psi_{\cal SE} \rangle \right)={\cal P}\left(|s_2\rangle {\Big |} |\psi_{\cal SE} \rangle \right)=1/2$, in agreement with the Born rule for the particular state $ |\psi_{\cal SE} \rangle$. This derivation is readily generalized to account for unequally weighted superpositions, as discussed in the Appendix A.  A number of further approaches to explaining the Born rule have since been proposed, that depend upon some or all of the Premises I-III introduced above, including most recently that of Carroll and Sebens~\cite{carroll:14}.

It should be emphasized, however, that Premises I-III do not represent assumptions that must be added to the set of basic foundational axioms upon which quantum theory is built. Rather, these premises can be {\it deduced} from more fundamental assumptions regarding the unitarity and linearity of time evolution in quantum mechanics, the combination of quantum systems by tensor product, and so forth (see Appendix B)~\cite{zurek:05}. Hence, the existing body of theoretical work supporting the preceding derivation, along with any associated experimental demonstrations of the validity of Premises I-III, ultimately serve to \textit{reduce} the number of axioms required to produce quantum mechanics by eliminating the need for the Born rule postulate without introducing additional complications to the theory. 

\begin{figure}[t]
	\centering
	\includegraphics[width=0.9\columnwidth]{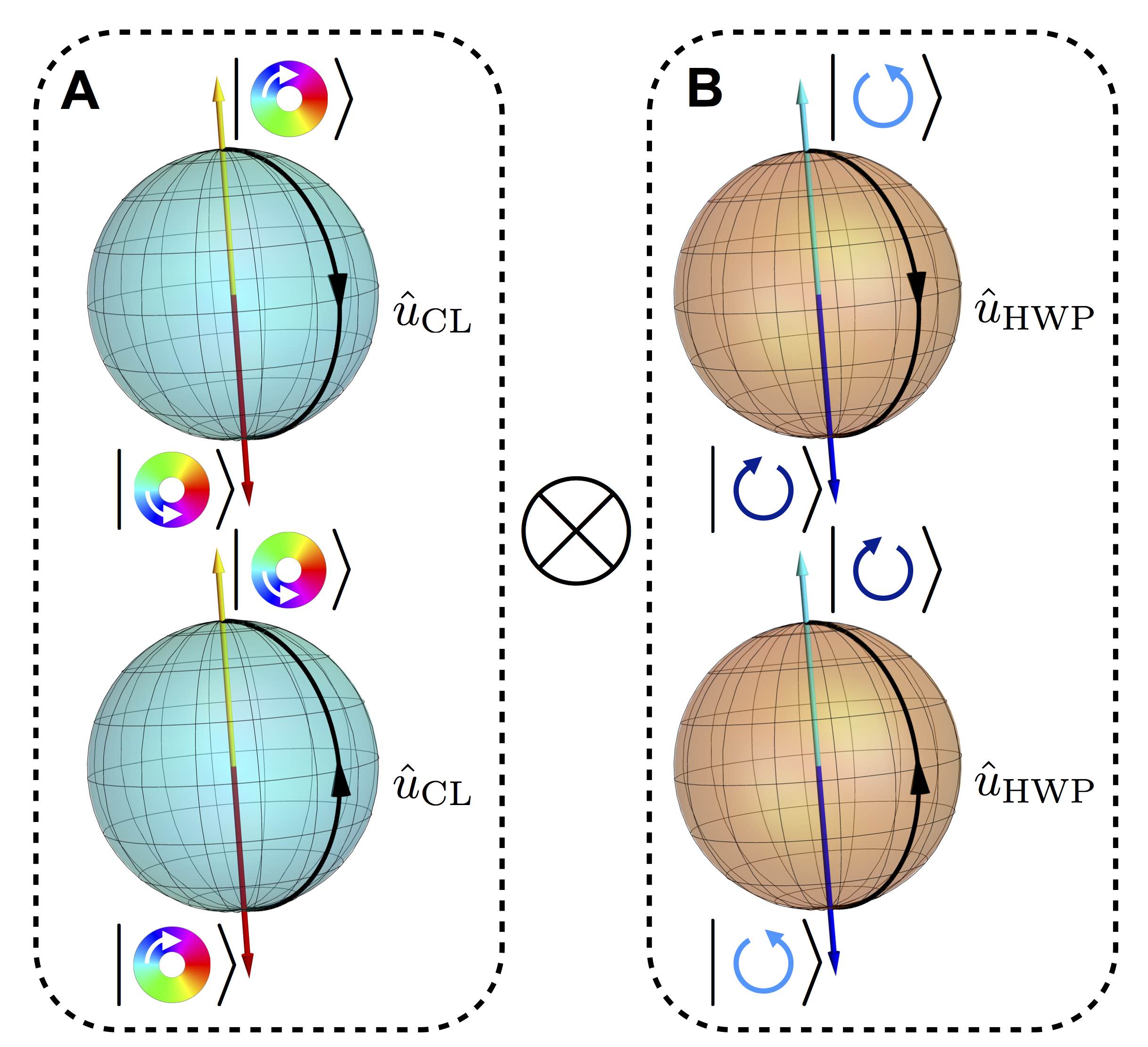}
	\caption{{\bf Effects of OAM and SAM swaps visualized on the Poincar\'e (Bloch) sphere.} {\bf (A)} Poincar\'e sphere representations of the effect of applying an OAM swap operator to the $|+1\rangle$ and $|-1\rangle$ OAM states~\cite{padgett:99}. The OAM swap operator $\hat{u}_\mathrm{CL}$ is physically realized by a $\pi/2$-cylindrical lens mode converter in Experiment 2 (see Appendix E). For both the $|+1\rangle$ and $|-1\rangle$ states, application of $\hat{u}_\mathrm{CL}$ results in a pole-to-pole rotation of the OAM state. {\bf (B)} Polarization state rotation resulting from the application of a SAM swap operator $\hat{u}_\mathrm{HWP}$, which, in both experiments, was realized physically by a half-wave plate. A one-to-one mapping exists between the $\pi$-rotation induced in the OAM state due to the cylindrical lenses, and the $\pi$-rotation induced in the SAM state by the half-wave plate.}
	\label{fig:fig1}
\end{figure}

In order to justify, and to confer physical validity to the condensed BPR derivation presented above, it is necessary to subject Premises I-III to comprehensive experimental verification. We carried out such an investigation using a bipartite quantum state in which the \qo{system} and \qo{environment} were identified with two internal photonic degrees of freedom, namely the photon spin angular momentum (SAM) and orbital angular momentum (OAM). We made use of OAM and SAM swapping operators, realized by different unitary optical components, as shown in Fig.~\ref{fig:fig1}.

%
\section{Experiment and Results}

\begin{figure}[t]
	\centering
	\includegraphics[width=0.9\columnwidth]{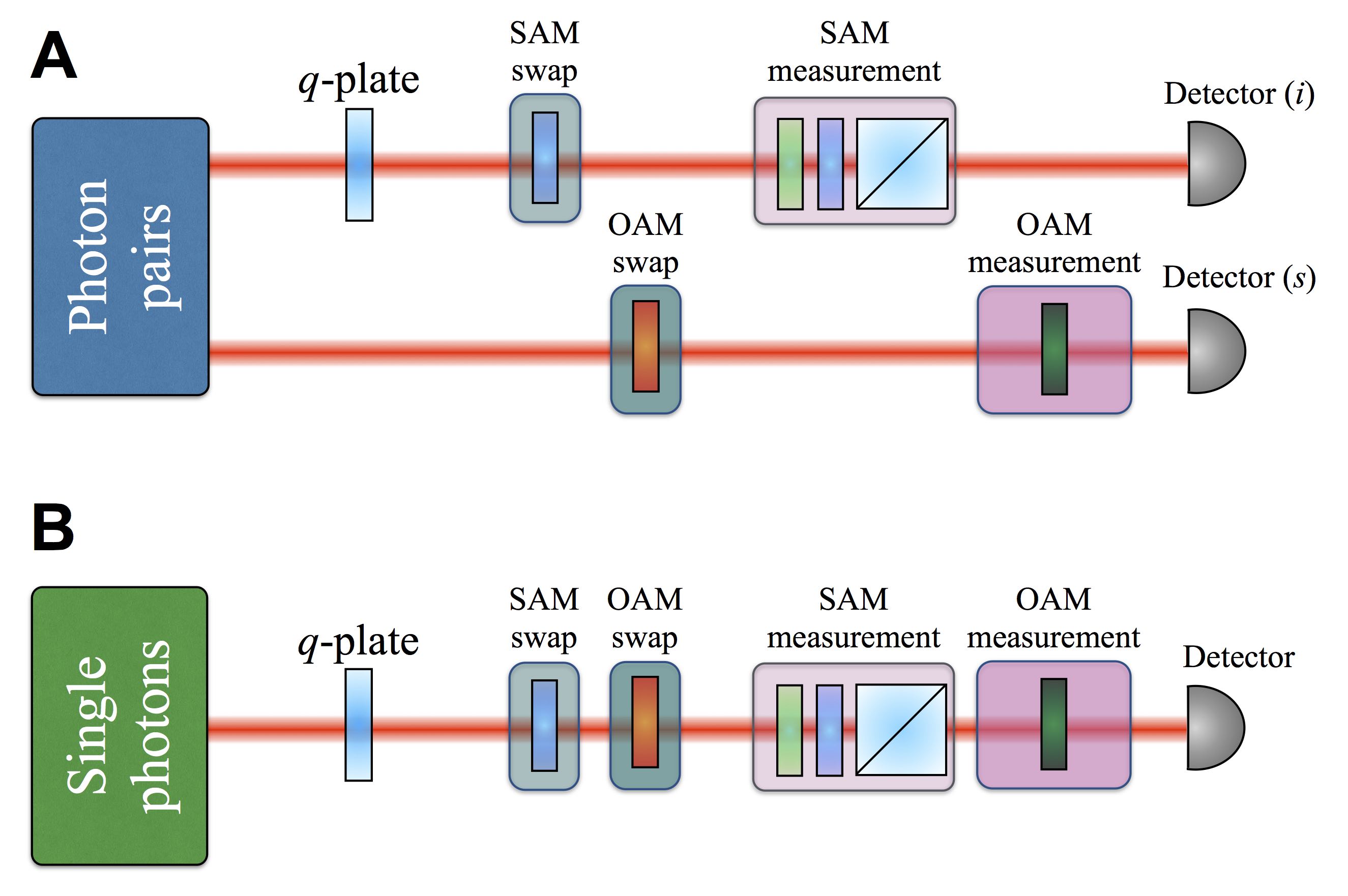}
	\caption{{\bf Experimental setups.} {\bf (A)} Experiment 1 setup testing Premises I-III in the nonlocal regime. An entangled photon pair is generated by spontaneous parametric down conversion and a \textit{q}-plate placed in the path of the idler. SAM and OAM swap operators, respectively realized by a half-wave plate and coordinate change, are placed in the path of the idler and signal. OAM tomography is executed using a spatial light modulator (SLM), while polarization tomography is achieved using a combination of waveplates and a polarizing beam splitter (WP/PBS). {\bf (B)} Experiment 2 setup, testing the validity of Premises I-III in the local regime. The OAM and SAM of a heralded single photon are entangled by a \textit{q}-plate, following which SAM and OAM swap operators, respectively consisting of a half-wave plate and a pair of cylindrical lenses, are applied to the state. Polarization and OAM tomography are achieved using a WP/PBS and a SLM, respectively.}
\label{fig:fig2}
\end{figure}

\begin{figure*}[t]
	\centering
	\includegraphics[width=\columnwidth]{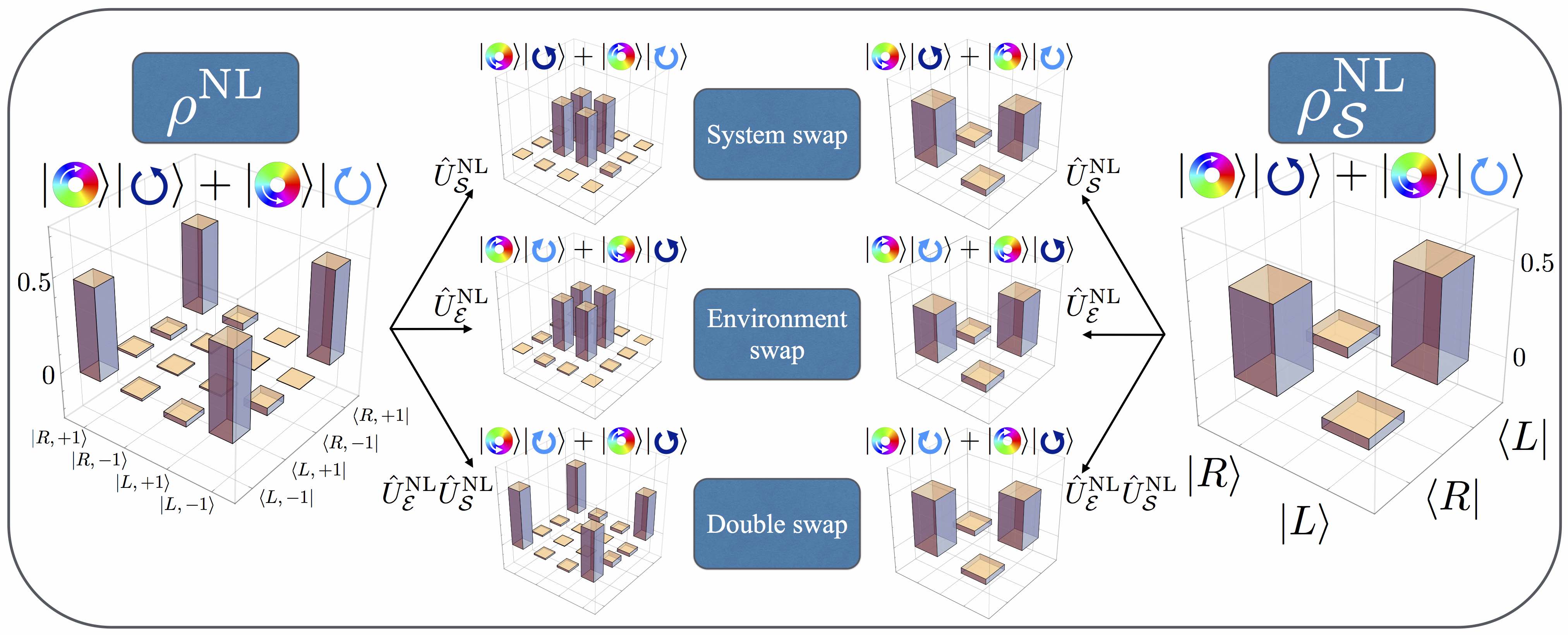}
	\caption{{\bf Experimental results.} Density matrices reconstructed from tomographic data obtained during the nonlocal test of Premises I-III carried out in Experiment 1. The real parts of the full density matrices $\rho^\mathrm{NL}$ and reduced system density matrices $\rho^\mathrm{NL}_{\cal S}$ obtained for the original, system (OAM)-swapped, environment (SAM)-swapped and twice-swapped states are displayed, along with \qo{wavefunction} representations of the states, showing the entanglement between the OAM and SAM degrees of freedom. The original and twice-swapped full density matrices are found to be highly similar, in agreement with Premise I, while the reduced density matrices are found to be largely indistinguishable, regardless of the application of either or both swapping operators, in agreement with Premise II. The imaginary parts of the density matrices plotted in the figure are displayed in the Appendix D.}
	\label{fig:fig3}
\end{figure*}

Our investigation is divided into two experiments. In Experiment 1, we generated an entangled photon pair by spontaneous parametric down-conversion, and used a set of linear optics, including a photonic \textit{q}-plate~\cite{marrucci:06}, to  produce a nonlocal state $|\psi_{\cal SE}^\mathrm{NL}\rangle$ in which the OAM degree of freedom of the signal $s$ photon was entangled with the polarization degree of freedom of the idler $i$. The resulting state takes the form $|\psi_{\cal SE}^\mathrm{NL}\rangle=1/\sqrt{2}\left(|R\rangle_i |+1\rangle_s + |L\rangle_i |-1\rangle_s\right)$, where we take $|R\rangle_i$ $\left(|L\rangle_i\right)$ to represent the right- (left-) handed circular polarization state of the idler photon, and $|+1\rangle_s$ $\left(|-1\rangle_s\right)$ the state of the corresponding signal photon carrying $+1$ $(-1)$ unit of OAM~\cite{karimi:10}. The experimental setup is shown in Fig.~\ref{fig:fig2}{\bf A}. We carried out tomography on the full system-environment (OAM-polarization) space and represented the generated state with a density matrix $\rho^\mathrm{NL}$. We then applied a first swap operator, $\hat{u}_{\cal S}^\mathrm{NL}=|R\rangle_i \langle L |_i + |L\rangle_i \langle R |_i$~{\cite{footnote4}}, to the state by placing a half-wave plate in the path of the idler photon, and performed tomography on the full system-environment space. Next, we applied an OAM-swap operator $\hat{u}_{\cal E}^\mathrm{NL}=|+1\rangle_s\langle -1 |_s + |-1\rangle_s \langle +1|_s$, by carrying out tomography once more on this new state, in an OAM basis that was the mirror-image complement of that used for tomography of the original, unswapped state. Finally, we applied both swap operators simultaneously, and carried out a last tomographic analysis, resulting in a density matrix representing the twice-swapped state. The corresponding reconstructed density matrices are plotted in Fig.~\ref{fig:fig3}.

We note that the similarity between any two quantum states is conventionally quantified by a fidelity computed from the direct comparison of the states' tomographically reconstructed density matrices. Unfortunately, such an analysis would be inappropriate in this case, given that quantum state tomography is a procedure which, itself, inherently assumes the BPR to be correct. The use of such a comparison strategy would therefore be subject to a charge of circularity~\cite{vermeyden:15}. Consequently, the quantum states generated here must be compared using an alternative figure of merit that does not assume the \textit{a priori} correctness of the Born rule, although we nonetheless display tomographically reconstructed density matrices, in the interest of providing readers with a familiar perspective on the generated states. An appropriate figure of merit for our present purpose is provided by the Bhattacharyya coefficient, a parameter that quantifies the extent of the similarity between two general probability distributions. In our case, the probability distributions in question correspond to the probability of a photon detection event for each of the $36$ projective measurements made to complete the tomographic reconstruction of each of the quantum states of interest. The Bhattacharyya coefficient, $B({\cal P}_1,{\cal P}_2)$, associated with two probability distributions ${\cal P}_1(i)$ and ${\cal P}_2(i)$, where $i$, in this case, denotes one of the $N=36$ projective measurements made on each state, is defined as $B({\cal P}_1,{\cal P}_2)=\sum_{i=1}^{N}\sqrt{{\cal P}_1(i){\cal P}_2(i)}$. In principle, $B({\cal P}_1,{\cal P}_2)$ can take on any value between $0$ and $1$, with the former case indicating maximal \textit{dissimilarity} between the probability distributions ${\cal P}_1(i)$ and ${\cal P}_2(i)$, and the latter their equality. In this investigation, we approximate the probability distributions ${\cal P}_1$ and ${\cal P}_2$ by the number of photon detection events registered for a particular state, and a particular arrangement of tomographic instruments, normalized by the total number of counts registered throughout the experiment for that particular state (hence, we determine the \textit{frequencies} of the associated detection events). The correspondence between the original state $|\psi_{\cal SE}^\mathrm{NL}\rangle$ and the twice-swapped state $\hat{U}_{\cal E}^\mathrm{NL} \hat{U}_{\cal S}^\mathrm{NL} |\psi_{\cal SE}^\mathrm{NL}\rangle$, as measured by the corresponding Premise I Bhattacharyya coefficient was found to be $B^\mathrm{NL}_\mathrm{I}({\cal P}_{o},{\cal P}_{\cal SE})=0.9976\pm0.0058$, where ${\cal P}_{o}$ and ${\cal P}_{\cal SE}$ respectively denote the probability distributions for the original and twice-swapped states. This result demonstrates that the unswapped and twice-swapped states show strongly statistically similar behaviour, and suggests that envariant transformations can be physically realized for nonlocally entangled states, in agreement with Premise I. 

In addition to experimentally generating the full density matrices $\rho^\mathrm{NL}$, we tested Premise II for this nonlocally entangled state by directly measuring the system alone in order to produce the system (the signal photon OAM) density matrix, without monitoring the environment (the idler photon polarization). In order for Premise II to hold, the statistics of the system must be invariant under application of either or both of the swapping operators $\hat{U}_{\cal S}^\mathrm{NL}$ and $\hat{U}_{\cal E}^\mathrm{NL}$ to the overall state. As can be discerned from the reduced density matrices shown in Fig.~\ref{fig:fig3}, the state of the system, as represented by the density operator $\rho_{\cal S}^\mathrm{NL}$, was not found to change appreciably upon application of either swap operator, nor was it affected by the double-swap transformation, in complete agreement with Premise II. Respective Premise II Bhattacharyya coefficients of $B^\mathrm{NL}_\mathrm{II}({\cal P}_{o},{\cal P}_{\cal S})=0.9984\pm0.0083$, $B^\mathrm{NL}_\mathrm{II}({\cal P}_{o},{\cal P}_{\cal E})=0.9998\pm0.0082$ and $B^\mathrm{NL}_\mathrm{II}({\cal P}_{o},{\cal P}_{{\cal SE}})=0.9989\pm0.0082$ were determined by comparing the original/system-swapped, original/environment-swapped and original/twice-swapped tomography data. If the Born rule is not assumed, this result is somewhat remarkable, no analogous circumstance being possible classically; in the classical case, composite physical systems are constructed from the Cartesian product of their subsystems, so that non-trivial operations carried out on one subsystem can never be reversed by acting on another~\cite{zurek:05} when the overall state is known (\qo{pure}). Indeed, it is precisely the fact that Hilbert spaces of quantum systems combine by \textit{tensor} product that allows for entanglement and enables the restoration of the original system/environment state via the counterswap operation. 

In order to test Premise III, it is necessary to establish a link between the statistics of the system and environment. We did so by measuring the probability of obtaining a polarization state $|R\rangle_s$ $\left(|L\rangle_s\right)$, given that the corresponding OAM state $|+1\rangle_i$ $\left(|-1\rangle_i\right)$ is selected. We found that, for the unswapped state, the $|R\rangle$ polarization state was detected along with the $|+1\rangle$ and $|-1\rangle$ OAM states with respective frequencies of $99.3\pm0.3\%$, and $0.7\pm0.3\%$. Conversely, the $|L\rangle$ polarization state was accompanied by the $|-1\rangle$ and $|+1\rangle$ OAM states with respective frequencies of $98.5\pm0.3\%$ and $1.5\pm0.4\%$. Both results suggest Premise III to be valid, when possible systematic errors, including manufacturing defects in the \textit{q}-plate, are considered.

%

At this stage, a subtle, but important point should be made regarding the BPR derivation introduced above: as was observed at the outset of the initial BPR derivation, Premise II could be understood to follow directly from causality. Indeed, if swapping environment states produced a change that could be detected by a system-only observer, violations of causality would ensue when the system and environment were spacelike separated. However, an appeal to relativistic causality is certainly undesirable if one wishes to derive the Born rule from purely quantum mechanical arguments, as is the case here. Premise II is, in fact, a more general statement that should, in principle, apply even to locally entangled degrees of freedom, otherwise known as local non-separable states~\cite{footnote:03}. Motivated by this observation, we test the local versions of Premises I-III in a second experiment, which decouples the constraint of envariance from that of causality.
%
\begin{figure*}[t]
	\centering
	\includegraphics[width=\columnwidth]{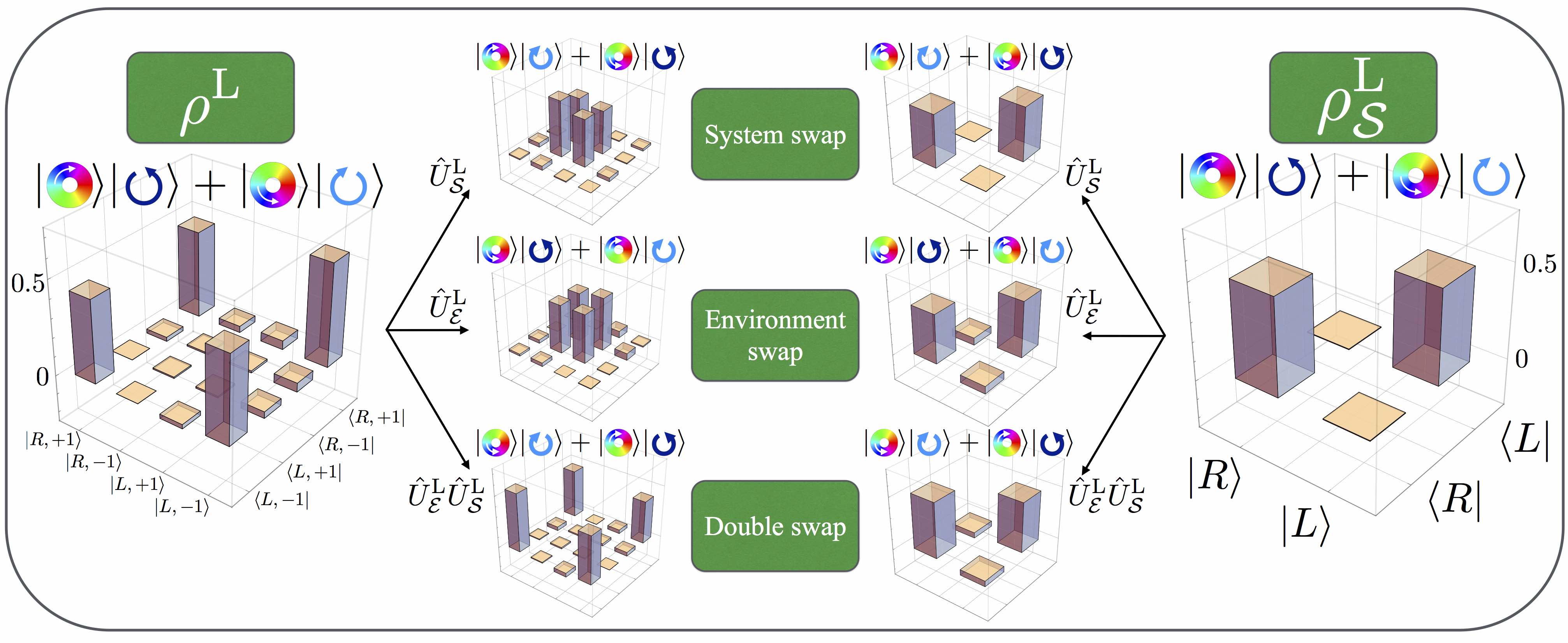}
	\caption{{\bf Experimental results.} Reconstructed density matrices obtained from quantum state tomography carried out during the local test of Premises I-III carried out in Experiment 2. The real parts of the full and reduced density matrices, $\rho^\mathrm{L}$ and $\rho^\mathrm{L}_{\cal S}$, obtained for the original, system (SAM)-swapped, environment (OAM)-swapped and twice-swapped states are displayed, along with \qo{wavefunction} representations of the states. The original and twice-swapped full density matrices are found to be highly similar, in agreement with Premise I, while the reduced density matrices are found to be largely indistinguishable, regardless of the application of either or both swapping operators, in agreement with Premise II. The imaginary parts of the density matrices plotted in the figure are displayed in the Appendix D.}
	\label{fig:fig4}
\end{figure*}
In Experiment 2, we used a heralded single-photon source to generate a single-photon nonseparable state of the form $|\psi_{\cal SE}^\mathrm{L}\rangle= 1/\sqrt{2} \left(|R\rangle |+1\rangle + |L\rangle |-1\rangle \right)$, in which the system is identified with the photon's polarization degree of freedom, and the environment with its OAM degree of freedom. The corresponding experimental setup is shown in Fig.~\ref{fig:fig2}{\bf B}. Tomographic analyses similar to those performed during Experiment 1 were carried out, in order to characterize the original, polarization-swapped (system-swapped), OAM-swapped (environment-swapped), and twice-swapped states, which are represented by the density matrices in Fig.~\ref{fig:fig4}. The polarization-swapped state was generated by applying a unitary operator $\hat{u}_{\cal S}^\mathrm{L}=|R\rangle \langle L|+|L\rangle \langle R|$ to the original state, using a half-wave plate, while the OAM-swapped state was obtained by applying a unitary operator $\hat{u}_{\cal E}^\mathrm{L}=|+1\rangle\langle-1| + |-1\rangle\langle+1|$ to the state, using a $\pi/2$-cylindrical lens mode converter~\cite{allen:92}, as already discussed in Fig.~\ref{fig:fig1}. The correspondence between the density matrices obtained for the original and twice-swapped states, along with the associated Premise I Bhachattaryya coefficient $B^\mathrm{L}_\mathrm{I}({\cal P}_{o},{\cal P}_{{\cal SE}})=0.9915\pm0.0041$ suggests that Premise I is satisfied even in the case of mutually local unitary transformations, to within $0.85\%$. 

Next, we tested Premise II for the locally entangled state $|\psi_{\cal SE}^\mathrm{L}\rangle$. As before, Premise II requires that no detectable changes occur in the system (polarization) space when either or both of the transformations $\hat{U_{\cal S}^\mathrm{L}}$ and $\hat{U_{\cal E}^\mathrm{L}}$ are applied to the state. In some sense, this is a more surprising consequence of envariance than that demonstrated in Experiment 1: here, a swap executed in one degree of freedom of a particular photon must not affect another degree of freedom possessed by that same photon. No appeal to relativistic causality can explain this phenomenon; rather, it is best understood with direct reference to envariance~\cite{footnote:04}. We find once again that the premises called upon by the BPR derivation discussed here are indeed verified in practice, the once-swapped and twice-swapped states showing no detectable disagreement, as evidenced by a comparison of the reduced density matrices by which they are represented in Fig.~\ref{fig:fig4}, and the corresponding Premise II Bhattacharyya coefficients $B^{L}_\mathrm{II}({\cal P}_{o},{\cal P}_{\cal S})=0.9995\pm0.0024$, $B^{L}_\mathrm{II}({\cal P}_{o},{\cal P}_{\cal E})=0.9998\pm0.0021$ and $B^{L}_\mathrm{II}({\cal P}_{o},{\cal P}_{{\cal SE}})=0.9994\pm0.0024$. 

Finally, we also tested the experimental validity of Premise III for this local state. We found that, for the unswapped state, the probabilities of registering a polarization state $|R\rangle$, given respective OAM states $|+1\rangle$ and $|-1\rangle$ were $98.1\pm0.3\%$ and $1.9\pm0.3\%$. In turn, we found that the $|L\rangle$ polarization state was accompanied by the $|-1\rangle$ and $|+1\rangle$ OAM states with respective frequencies of $98.4\pm0.3\%$ and $1.6\pm0.3\%$, showing close correspondence with Premise III. {We close with a brief comment to place this work in context, particularly with respect to the results presented in the investigation~\cite{vermeyden:15}. While reference~\cite{vermeyden:15} and the present study both reported an experimental investigation of envariance, the two projects differed considerably in scope and aim. In particular, \cite{vermeyden:15} presented a rigorous investigation of what we here have referred to as Premise I, in the case of a nonlocal quantum state. This commonality notwithstanding, the present work is the first to report an experimental test of Premises II and III, and the first to explore the envariance argument experimentally in a purely local system. As we have discussed, this latter point is important: while many of the effects of envariance can be explained by appeals to causality for nonlocal systems, envariance can be fully distinguished from locality only if one carries out experiments on a purely local state of the form used in the present work.}

\section{Conclusion}

In summary, we have demonstrated experimentally the validity of three key premises involved in a decoherence-motivated derivation of the Born probability rule. Specifically, we have shown, without appealing to the BPR, that an experiment sensitive to one part of a bipartite, entangled quantum system in a particular state $|\psi_{\cal SE}\rangle$ cannot detect a \qo{state-swapping} action carried out over an entangled but unmonitored degree of freedom of that same state. More remarkably still, we have demonstrated that this same experiment cannot detect a swap carried out on the \textit{very degree of freedom being monitored}, for the state under consideration. We note that, in the absence of the Born rule postulate, such a finding is nontrivial, and highly counterintuitive. We have also extended a previous demonstration of the symmetry of envariance to a domain in which an appeal to causality cannot be made to explain experimental observations, and have thereby shown the more general validity of this principle, as invoked in the original BPR argument. {Our experiment can therefore be understood to establish a lower bound on the extent to which nature observes the Born probability rule. This lower bound is set by the range of uncertainties associated with the Bhattacharyya coefficients and outcome frequencies obtained in our experiments. Consequently, we have provided strong empirical support for the envariance-based derivation of the Born rule, and have taken an important step toward explaining the emergence of classical reality from the mathematical formalism of quantum mechanics.}

\section{Acknowledgments}
J.H., F.B., R.W.B. and E.K. acknowledge the support of the Canada Excellence Research Chairs (CERC) Program. J.H. acknowledges the support of the Vanier Canada Graduate Scholarships Program. F.B. acknowledges the support of the Natural Sciences and Engineering Research Council's Canada Graduate Scholarships program. W.H.Z was supported by DoE under the Los Alamos LDRD program and, in part, by the Foundational Questions Institute Grant No. 2015-144057 on \qo{Physics of What Happens}. E.K. acknowledges the support of the Canada Research Chairs (CRC) program. The authors would like to thank Lorenzo Marrucci for the $q$-plate preparation, as well as Harjaspreet Mand, Alicia Sit, and Sarina Cotroneo for their assistance with the experiment.

\appendix
\section{Generalization to arbitrary coefficients:}
In what follows, we present a generalization of the Born rule derivation discussed in the main text, to the case of arbitrary coefficients in the state Schmidt representation. Ours is a simplified version of this generalization~\cite{zurek:03a, zurek:05, zurek:14}, which followed the original BPR argument. In this approach, we consider an \qo{imbalanced} state of the form
\begin{eqnarray}\label{eq:genpsi}
	|\psi\rangle=\left(\sqrt{2}|s_0\rangle|\varepsilon_0\rangle+|s_2\rangle|\varepsilon_2\rangle\right)/\sqrt{3}, 
\end{eqnarray}
generated as a result of a pre-measurement by the environment on the system, where $\langle s_i|s_j\rangle=\langle \varepsilon_i|\varepsilon_j\rangle=\delta_{i,j}$, and the system and environment are respectively spanned by the orthonormal kets $\{|s_1\rangle,|s_2\rangle\}$ and $\{|\varepsilon_0\rangle,|\varepsilon_1\rangle,|\varepsilon_2\rangle\}$. We note at the outset of this derivation that the state $|\psi\rangle$ is clearly \textit{not} envariant under a swap of its system and environment state labels, precisely due to the presence of the $\sqrt{2}$ coefficient present in Eq.~(\ref{eq:genpsi}). The environment must therefore be assumed to possess at least a dimensionality of 3, but we note that this requirement is satisfied in the vast majority of real-world cases, in which the environment possesses a number of degrees of freedom far greater than that possessed by the system. We are then justified in defining a new pair of environment kets such that $|\pm\rangle=\left(|\varepsilon_0\rangle\pm|\varepsilon_1\rangle\right)/\sqrt{2}$. Under these conditions, we can rewrite the original state as follows:
\begin{eqnarray}\label{eq:genpsi}
	|\psi\rangle=\left(\sqrt{2}\left(|s_0\rangle|+\rangle+|s_0\rangle|-\rangle\right)/\sqrt{2}+|s_2\rangle|\varepsilon_2\rangle\right)/\sqrt{3}.
\end{eqnarray}
The $\sqrt{2}$ terms cancel, whence we find that 
\begin{eqnarray}\label{eq:genpsi}
	|\psi\rangle=\left(|s_0\rangle|+\rangle+|s_0\rangle|-\rangle+|s_2\rangle|\varepsilon_2\rangle\right)/\sqrt{3}.
\end{eqnarray}
The present strategy, which might be referred to as a \qo{fine-graining} of the state $|\psi\rangle$, has allowed us to write a representation of $|\psi\rangle$ in which the state is expressed as a sum of equally weighted kets, at the expense of enlarging the Hilbert space over which it is defined. At this point, it may be tempting to invoke the envariance-based argument presented in the main text to argue for the Born rule in a more general form. However, this strategy is ill-fated, as we now will show. If, for example, we introduce the necessarily non-unitary swapping operator pair $\hat{O}_{\cal S}=|s_0\rangle\langle s_2| + |s_2\rangle\langle s_0|$ and $\hat{O}_{\cal E}=|+\rangle\langle \varepsilon_2| + |\varepsilon_2\rangle\langle +|$, with the aim of establishing the equiprobability of the $|s_0\rangle|+\rangle$ and $|s_2\rangle|\varepsilon_2\rangle$ states, we find that
\begin{eqnarray}\label{eq:genpsi}
	\hat{O}_{\cal E}\hat{O}_{\cal S}|\psi\rangle=\left(|s_2\rangle|\varepsilon_2\rangle+|s_0\rangle|+\rangle\right)/\sqrt{3}\neq|\psi\rangle,
\end{eqnarray}
so that arguments based upon envariance may not be invoked. However, this difficulty is overcome by the introduction of a \textit{second} environment, spanned by the orthonormal states $\{|e_+\rangle,|e_-\rangle,|e_2\rangle\}$, which is made to interact with the first environment, resulting in a second pre-measurement interaction of the form
\begin{eqnarray}\label{eq:genpsi}
	|\psi\rangle&=&1/\sqrt{3}\left(|s_0\rangle|+\rangle+|s_0\rangle|-\rangle+|s_2\rangle|\varepsilon_2\rangle\right)|e_+\rangle \\\nonumber &\rightarrow&\left(|s_0\rangle|+\rangle|e_+\rangle+|s_0\rangle|-\rangle|e_- \rangle+|s_2\rangle|\varepsilon_2 \rangle | e_2\rangle\right)/\sqrt{3}=|\psi'\rangle.
\end{eqnarray}
We are now in a position to set the scene for the standard envariance argument in this context. We begin by defining the composite kets $|s_0\rangle|+\rangle=|s_0^+\rangle$, $|s_0\rangle|-\rangle = |s_0^-\rangle$ and $|s_2\rangle|\varepsilon_2\rangle = |s_2^2\rangle$, so that we may now write
\begin{eqnarray}\label{eq:genpsi}
	|\psi'\rangle=1/\sqrt{3}\left(|s_0^+\rangle|e_+\rangle+|s_0^-\rangle|e_-\rangle+|s_2^2\rangle|e_2\rangle\right).
\end{eqnarray}
We next define the unitary swapping operators
\begin{eqnarray}\label{eq:genpsi}
	\hat{U}^{+,-}_{\cal SE}&= |s_0^+\rangle\langle s_0^-| + |s_0^-\rangle\langle s_0^+| + |s_2^2\rangle \langle s_2^2| \\
	\hat{U}^{+,-}_{E}&= |e_+\rangle \langle e_-| + |e_-\rangle \langle e_+| + |e_2\rangle \langle e_2| \\
	\hat{U}^{-,2}_{\cal SE}&=|s_0^+\rangle \langle s_0^+| + |s_0^-\rangle \langle s_2^2| + |s_2^2\rangle \langle s_0^-|  \\
	\hat{U}^{-,2}_{E}&= |e_+\rangle \langle e_+| + |e_-\rangle \langle e_2| + |e_2\rangle \langle e_-|.
\end{eqnarray}
We note that these operators satisfy the envariance relations $\hat{U}^{+,-}_{E}\hat{U}^{+,-}_{\cal SE}|\psi'\rangle=|\psi'\rangle$ and $\hat{U}^{-,2}_{E}\hat{U}^{-,2}_{\cal SE}|\psi'\rangle=|\psi'\rangle$. As a result, one concludes from the standard envariance argument that ${\cal P}\left(|e_+\rangle \Big{|} |\psi'\rangle\right)={\cal P}\left(|e_-\rangle \Big{|} |\psi'\rangle\right)={\cal P}\left(|e_2\rangle \Big{|} |\psi'\rangle\right)$, and by Premise III, that ${\cal P}\left(|s_0\rangle \Big{|} |\psi'\rangle\right)=2 {\cal P}\left(|s_2\rangle \Big{|} |\psi'\rangle\right)=2/3$, as expected from the Born rule. A fully general treatment of the case of arbitrary coefficients was discussed in the original Born rule derivation, as well as in subsequent work in the field~\cite{zurek:03a,zurek:11}. 

\section{Additional remarks on Premises I-III:}
In the interest of clarity, we now turn to a brief discussion of terminology. Throughout the main text, we refer to a set of quantum mechanical principles (\qo{Premises I-III}) from which the Born probability rule can be derived. We wish to emphasize here what has already been argued in the text: these premises should not be understood to represent \textit{assumptions} that must be added to the set of basic \qo{no-collapse} axioms of quantum mechanics. Instead, these are propositions regarding the behaviour of quantum systems that may be derived from standard foundational axioms upon which quantum theory is generally constructed, absent the Born rule postulate. The premises arise from a set of assumptions about the quantum Universe (the so-called \qo{quantum credo}), which are already contained as subsets of the collection of foundational axioms required to formulate no-collapse quantum mechanics~\cite{zurek:14,zurek:05}:

\begin{enumerate}
\item Quantum states correspond to vectors in a Hilbert space. 
\item Quantum states undergo unitary evolution under the Schr\"odinger equation, $i\hbar\partial_t|\psi\rangle=\hat{H}|\psi\rangle$, where $\hat{H}$ is Hermitian.
\item Composite quantum states can be expressed as superpositions of the form $\sum_{i,j}c_{i,j}|s_i\rangle|\varepsilon_j\rangle$.
\item An immediately repeated measurement will yield the same outcome as its preceding measurement.
\end{enumerate}

Our experimental verification therefore serves a very specific purpose: to show that the physical world indeed abides by the quantum mathematics that lead one to the BPR derivation presented in the main text. It is therefore the gap between the mathematical and physical descriptions of the universe that we have sought to bridge.
\begin{figure*}[t]
	\centering
	\includegraphics[width=\columnwidth]{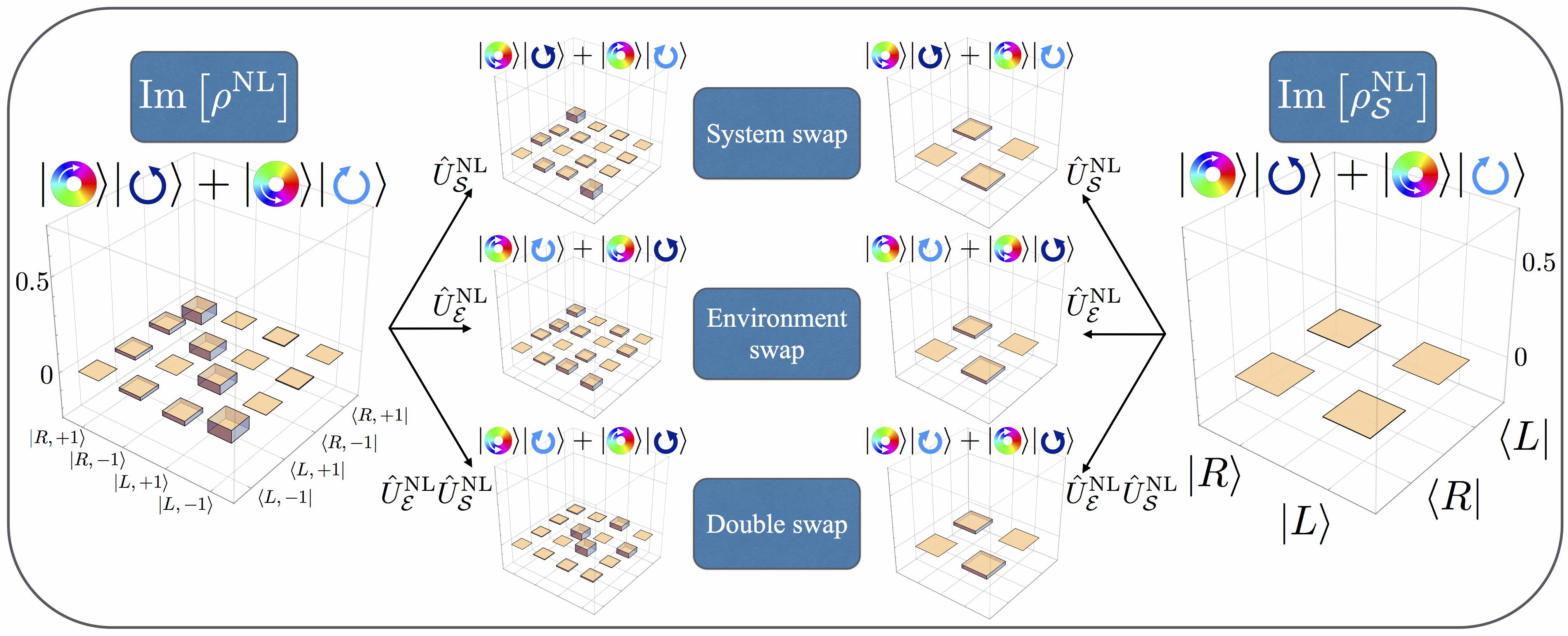}
	\caption{{\bf Imaginary density matrices obtained from Experiment 1 tomographic data.} Imaginary parts of the reconstructed density matrices obtained from quantum state tomography carried out during the nonlocal test of Premises I-III carried out in Experiment 1. The imaginary parts of the full and reduced density matrices, $\rho^{L}$ and $\rho^{L}_{\cal S}$, obtained for the original, system (SAM)-swapped, environment (OAM)-swapped and twice-swapped states are displayed, along with wavefunction representations of the states. The original and twice-swapped full density matrices are found to be highly similar, in agreement with Premise I, while the reduced density matrices are found to be largely indistinguishable, regardless of the application of either or both swapping operators, in agreement with Premise II.}
	\label{fig:fig5}
\end{figure*}
%

\section{Untested \qo{candidate} Premises:}
\noindent {\bf 1. The eigenvalue-eigenstate link:} One might be led to wonder whether the set of three premises discussed in the main text is, indeed, exhaustive. Specifically, one might be inclined to suggest that a fourth premise, that of the \qo{eigenvalue-eigenstate link} (EEL), is also required, at least implicitly, by the BPR derivation presented in the text. The EEL postulates that a particular measurement outcome indicates the presence of the corresponding quantum state. According to the EEL, a photon detection event registered for a particular arrangement of the spatial light modulator and waveplates indicates the presence of a photon with the corresponding polarization and OAM. Despite its apparent relevance to the envariance-based BPR derivation, the eigenvalue-eigenstate link is, in fact, already captured as a special case of Premise III. In particular, consider the following pre-measurement interaction between a system space, spanned by \{$|s_1\rangle,|s_2\rangle$\}, and an ancillary space, spanned by \{$|a_1\rangle,|a_2\rangle$\}, where $\langle s_i|s_j\rangle=\langle a_i|a_j\rangle=\delta_{ij}$:
\begin{eqnarray}\label{eq:PreMeasurement2}
	\frac{1}{\sqrt{2}}\left(|s_1\rangle + |s_2\rangle\right) |a_1\rangle \rightarrow \frac{1}{\sqrt{2}}\left(|s_1\rangle|a_1\rangle + |s_2\rangle|a_2 \rangle \right).
\end{eqnarray}
We note that the correlation established between system and ancilla states in Eq.~(\ref{eq:PreMeasurement2}) can be interpreted as having imprinted upon the ancilla kets a \textit{de facto} record of the corresponding state of the system. This is precisely consistent with the ancilla's having perceived the measurement outcome associated with the particular system state with which it is partnered, and therefore with its having \qo{recorded} the pre-measurement. One consequence of this interaction is therefore the apparent emergence of an \qo{eigenvalue-eigenstate link}, so that the connection between measurement outcomes and measured states need not be postulated separately from Premise III. \\

\noindent {\bf 2. Null Schmidt coefficients:} A second candidate \qo{untested premise} so appeals to the intuition that it is easily overlooked. When represented in its Schmidt form, $|\psi\rangle = \sum_i c_i |s_i\rangle|\varepsilon_i\rangle$, a general bipartite quantum state contains kets $|s_k\rangle|\varepsilon_k\rangle$ for which the associated coefficients $c_k$ are equal to zero. Indeed, this is precisely the case for the state displayed in Eq.~(\ref{eq:PreMeasurement}), where, we have $c_1=c_2=1/\sqrt{2}$, $c_k=0$ for $k\geq3$. A careful reading of the derivation presented in the main text will reveal an additional tacit premise required by the BPR argument: states with null Schmidt coefficients were taken to be physically inadmissible, \textit{i.e.} $c_k=0$ was taken to indicate that the state $|s_k\rangle$ could never be observed upon measurement. This premise was called upon implicitly when it was suggested that the Born rule was satisfied by the relation ${\cal P}\left(|s_1\rangle {\Big |} |\psi_{\cal SE} \rangle \right)={\cal P}\left(|s_2\rangle {\Big |} |\psi_{\cal SE} \rangle \right)$, as it leads to the conclusion that ${\cal P}\left(|s_1\rangle {\Big |} |\psi_{\cal SE} \rangle \right)+{\cal P}\left(|s_2\rangle {\Big |} |\psi_{\cal SE} \rangle \right)=1$, from which the BPR follows. But to take the case $c_k=0$ to indicate the impossibility of observing outcome $|s_k\rangle$ is to assume, if only to a very limited degree, the Born rule \textit{a priori}. Should one conclude, then, that an additional premise ought to be tested in order to fully support the BPR derivation provided in the text?

Fortunately, the physical inadmissibility of states with null Schmidt coefficients has already been shown~\cite{zurek:08} to follow from the premises already tested during this experiment. In particular, arguments on envariance can be used to demonstrate that any envariantly swappable pair of states in the Schmidt representation of a bipartite quantum state must be equiprobable. For instance, if there exist $N$ terms in the Schmidt decomposition of a state $|\psi\rangle$ for which the associated prefactors $c_k=\alpha$, with $k=1,2,...,N$, one finds that
\begin{eqnarray}\label{eq:nullschmidt1}
	|\psi\rangle&=& \sum_i c_i |s_i\rangle|\varepsilon_i\rangle\\\nonumber 
	&=& \alpha\left(|s_1\rangle|\varepsilon_1\rangle + |s_2\rangle|\varepsilon_2\rangle + ... + |s_N\rangle|\varepsilon_N\rangle \right)+ \sum_{i=N+1}c_i |s_i\rangle|\varepsilon_i\rangle.
\end{eqnarray}

It follows from envariance arguments that ${\cal P}\left(|s_1\rangle \Big{|} |\psi\rangle\right)={\cal P}\left(|s_2\rangle \Big{|} |\psi\rangle\right)=...={\cal P}\left(|s_N\rangle \Big{|} |\psi\rangle\right)=p$, so that $P=Np$ is the probability of obtaining any one of the $N$ equally weighted states outside the sum in Eq.~(\ref{eq:nullschmidt1}). For the special case $\alpha=0$, we find that terms outside the sum in Eq.~(\ref{eq:nullschmidt1}) can be combined with impunity: for example, $0\left(|s_1\rangle|\varepsilon_1\rangle + |s_N\rangle|\varepsilon_N\rangle\right) = 0|s_{1'}\rangle|\varepsilon_{1'}\rangle$. If the state kets labelled $1$ and $N$ are combined in this way, one obtains
\begin{eqnarray}\label{eq:nullschmidt2}
	|\psi\rangle=0\left(|s_{1'}\rangle|\varepsilon_{1'}\rangle + |s_2\rangle|\varepsilon_2\rangle + ... + |s_{N-1}\rangle|\varepsilon_{N-1}\rangle \right)+ \sum_{i=N+1}c_i |s_i\rangle|\varepsilon_i\rangle.
\end{eqnarray}
The combination of $|s_1\rangle|\varepsilon_1\rangle$ and $|s_N\rangle|\varepsilon_N\rangle$ that leads to Eq.~(\ref{eq:nullschmidt2}) cannot have any physical consequences, however, as it merely involves a mathematical rearrangement of the state $|\psi\rangle$. Consequently, the total probability associated with the kets outside the summation must still be equal to the analogous probability in Eq.~(\ref{eq:nullschmidt2}), but there are now only $N-1$ such terms, all of which must be still be assigned an equal probability based upon envariance arguments. Hence, we must now have ${\cal P}\left(|s_{1'}\rangle \Big{|} |\psi\rangle\right)={\cal P}\left(|s_2\rangle \Big{|} |\psi\rangle\right)=...={\cal P}\left(|s_{N-1}\rangle \Big{|} |\psi\rangle\right)=p$, but $P=p(N-1)$. By equating the total probabilities $P$ assigned to the states outside the sums in Eqs.~(\ref{eq:nullschmidt1}) and (\ref{eq:nullschmidt2}), one finds that $pN=p(N-1)$, which can be satisfied only for $p=0$.
\begin{figure*}[t]
	\centering
	\includegraphics[width=\columnwidth]{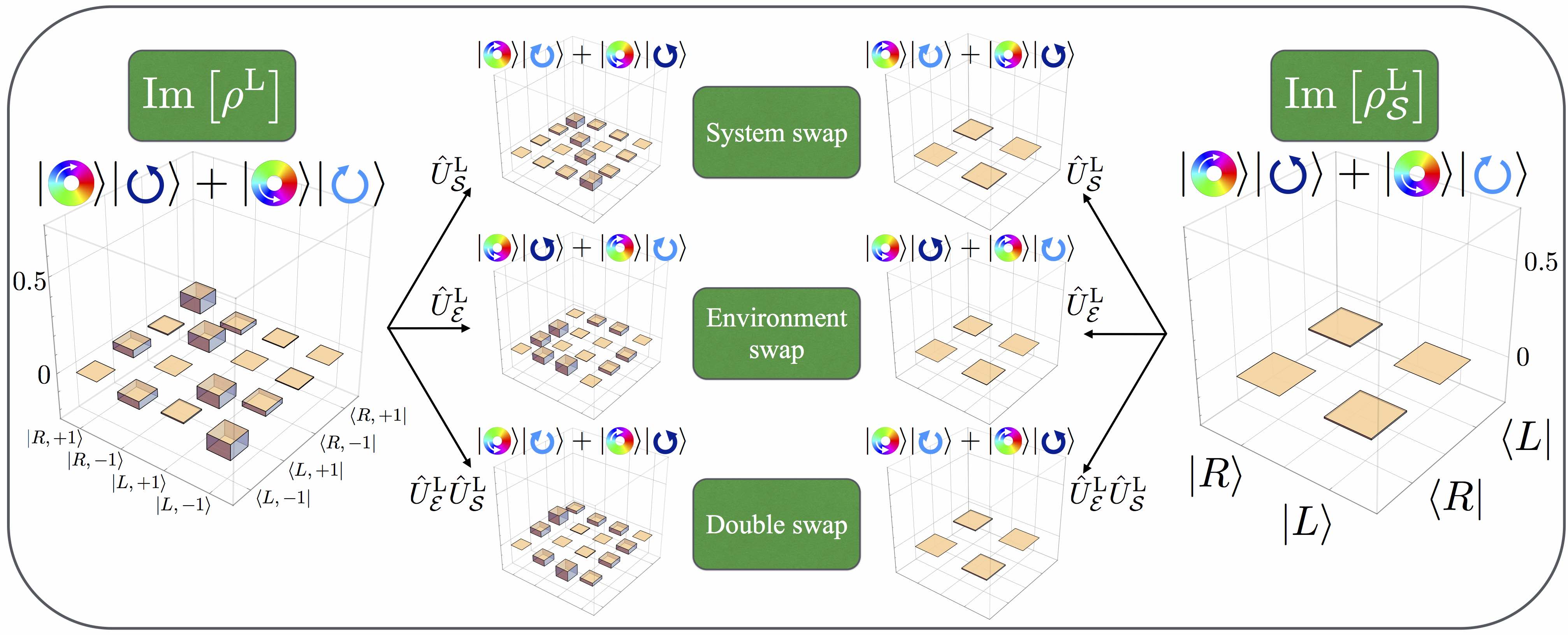}
	\caption{{\bf Imaginary density matrices obtained from Experiment 2 tomographic data.} Imaginary parts of the reconstructed density matrices obtained from quantum state tomography carried out during the local test of Premises I-III carried out in Experiment 2. The imaginary parts of the full and reduced density matrices, $\rho^{L}$ and $\rho^{L}_{\cal S}$, obtained for the original, system (SAM)-swapped, environment (OAM)-swapped and twice-swapped states are displayed, along with wavefunction representations of the states. The original and twice-swapped full density matrices are found to be highly similar, in agreement with Premise I, while the reduced density matrices are found to be largely indistinguishable, regardless of the application of either or both swapping operators, in agreement with Premise II.}
	\label{fig:fig6}
\end{figure*}

\section{Additional experimental data:}
In addition to the real part of the density matrices plotted in Fig.~\ref{fig:fig3} and Fig.~\ref{fig:fig4}, the tomographic data obtained throughout the experiment can, of course, be used to generate the imaginary parts of these same matrices. The imaginary density matrices corresponding to the data plotted in Fig.~\ref{fig:fig3} and Fig.~\ref{fig:fig4} are plotted in Fig.~\ref{fig:fig5} and Fig.~\ref{fig:fig6}, respectively. We note that the Bhattacharyya coefficients presented in the main text accounted for all raw tomographic data, and therefore reflect the combined information content required to produce both the real and imaginary density matrices obtained in Experiments 1 and 2. 

Further, although a fidelity-based comparison of the density matrices displayed in Figs.~\ref{fig:fig3}, \ref{fig:fig4}, \ref{fig:fig5} and \ref{fig:fig6} is inappropriate for the purpose of verifying Premises I-III, such an analysis may nonetheless be of interest to readers wishing for a more familiar comparison metric by which to assess the tomographic data. In Experiment 1, we found the relative fidelity of the original state $|\psi^\mathrm{NL}_{\cal SE}\rangle$ and twice-swapped state $\hat{U}^\mathrm{NL}_{\cal E}\hat{U}^\mathrm{NL}_{\cal S}|\psi^\mathrm{NL}_{\cal SE}\rangle$ to be $0.978 \pm 0.011$, while in Experiment 2, the analogous fidelity of the original state $|\psi^\mathrm{L}_{\cal SE}\rangle$ and twice-swapped state $\hat{U}^\mathrm{L}_{\cal E}\hat{U}^\mathrm{L}_{\cal S}|\psi^\mathrm{L}_{\cal SE}\rangle$ was found to be $0.940 \pm 0.011$. 

We also obtained relative fidelities of $0.9918 \pm 0.0028$, $0.9983 \pm 0.0017$ and $0.9949 \pm 0.0023$, respectively for the reduced system density matrices corresponding to the original/system-swapped, original/environment-swapped and original/twice-swapped cases in Experiment 1. In Experiment 2, the original/system-swapped, original/environment-swapped and original/twice-swapped density matrices were found to exhibit relative fidelities of $0.9998 \pm 0.0001$, $0.9951 \pm 0.0010$ and $0.9975 \pm 0.0007$. 

Finally, we determined the purities $p=\mathrm{Tr}[\rho_{\cal S}^2]$ associated with the reduced system density matrices $\rho_{\cal S}$ in both experiments. These quantities provide a valuable measure of the extent of the \qo{classicality} of the system, a purity of $1/2$ in this case indicating a perfectly mixed state exhibiting no coherence effects. The respective purities of the system were found to be $0.5100 \pm 0.0029$, $0.5145 \pm 0.0036$, $0.5145 \pm 0.0035$ and $0.5010 \pm 0.0029$ for the original, system-swapped, environment-swapped and twice-swapped reduced density matrices in Experiment 1, and $0.5002 \pm 0.0001$, $0.5003 \pm 0.0001$, $0.5031 \pm 0.0005$ and $0.5017 \pm 0.0004$ for the original, system-swapped, environment-swapped and twice-swapped reduced density matrices in Experiment 2.

\section{Experimental details:}
\noindent {\bf{Experiment 1:}} We use a 355~nm 150~mW quasi-continuous-wave laser to pump a Type I $\beta$-barium borate (BBO) nonlinear crystal, phase-matched for spontaneous parametric down conversion (SPDC). The signal and idler photons are split into two separate arms at a non-polarizing beams splitter. In one arm, we place a photonic \textit{q}-plate, along with a polarization tomography setup consisting of a quarter- and half-wave plate followed by a polarizing beam splitter. As needed, a polarization swapping operator, consisting of a half-wave plate, is placed between the \textit{q}-plate and the polarization tomography setup. In the second arm, we place a spatial light modulator (SLM) in the path of the beam, and use it to carry out OAM tomography, by means of a phase-flattening algorithm (see Tomography). The OAM swap operator is realized by carrying out tomography in a mirror-image basis to that used for tomography on the unswapped state. This is effectively equivalent to adding a mirror in the second arm, which corresponds to a substitution of $|+1\rangle$ for $|-1\rangle$, and vice-versa, and hence, an OAM swap of the required form, without affecting either photon's polarization state. By post-selecting only on the phase-flattened component of the beams emerging from the first and second arms, we produce the required state. The set of 36 tomographic measurements made on the combined OAM-polarization space can be used to reconstruct the full density matrices displayed in the text. 

Reduced tomography over the system (OAM) space was achieved by removing the polarization tomography setup from the first arm, therefore rendering the measurement insensitive to the polarization of the signal photon. The six measurements required for OAM tomography are carried out by displaying six different diffraction patterns on the SLM, and registering the corresponding coincidence counts obtained from the pair of single-photon detectors at the output of the setup. \\

\noindent The preparation and post-selection involved in measuring the unswapped state proceeds as follows: \\

\noindent 1. After SPDC, there will exist OAM-entangled photon pairs whose joint states are given by
\begin{eqnarray}\label{eq:genpsi}
	1/\sqrt{2}\left(|+1\rangle_s |-1\rangle_i + |-1\rangle_s |+1\rangle_i \right)|H\rangle_s|H\rangle_i.
\end{eqnarray} \\
2. When tuned, the \textit{q}-plate, for which $q=1/2$ converts a polarization state $|R\rangle$ ($|L\rangle$) into a state $|L\rangle$ ($|R\rangle$), while simultaneously decreasing (increasing) the OAM by one unit. Hence, following the $q$-plate, we are left with a state of the form
\begin{eqnarray}\label{eq:genpsi}
	1/2&{\bigg(}|0\rangle_s|-1\rangle_i|L\rangle_s + |0\rangle_s|+1\rangle_i|R\rangle_s \cr 
	&+ |+2\rangle_s|-1\rangle_i|L\rangle_s + |-2\rangle_s|+1\rangle_i|R\rangle_s {\bigg)} |H\rangle_i.
\end{eqnarray} \\
3. By post-selecting on the signal $|0\rangle_s$ state, and ignoring the idler polarization, one obtains a state in which the OAM of the idler is entangled with the polarization of the signal photon:
\begin{eqnarray}\label{eq:genpsi}
	1/\sqrt{2}\left(|-1\rangle_i|L\rangle_s + |+1\rangle_i|R\rangle_s \right),
\end{eqnarray}
as required.\\

\noindent {\bf{Experiment 2:}} We use a 355~nm 150~mW quasi-continuous-wave laser to pump a Type I $\beta$-barium borate (BBO) nonlinear crystal, phase-matched for spontaneous parametric down conversion (SPDC). The idler photon is used to herald the arrival of the signal, and the horizontally-polarized signal photon is passed through a photonic $q$-plate, for which $q=1/2$. Following the $q$-plate, the signal photon is made to pass through a $\pi/2$-cylindrical lens mode converter, which swaps the photon's OAM state, if desired. If a polarization swap operator is required, a half-wave plate can be placed in the beam path as well, but in either case, the beam is subsequently transmitted through a polarization tomography setup, consisting of a quarter-wave plate, half-wave plate, and polarizing beam splitter (PBS). Upon emerging from the PBS, the beam reaches a SLM, which can be used for OAM tomography, and coupled to a single-mode optical fiber (see: Tomography). The set of 36 tomographic measurements made on the combined OAM-polarization space can be used to reconstruct the full density matrices displayed in the text. 

In order to carry out reduced tomography over the system (polarization) space alone, we coupled the beam emerging from the polarization tomography setup directly to a multimode optical fiber, thereby circumventing the SLM, and tracing over the OAM degree of freedom. The multimode fiber effectively serves as a bucket detector in this context, as it couples the $|+1\rangle$ and $|-1\rangle$ OAM states with equal, nonzero, efficiency. \\

\noindent The preparation and post-selection involved in measuring the unswapped state proceeds as follows: \\

\noindent 1. The linearly polarized single-photon state entering the experimental apparatus takes the form $|H\rangle|0\rangle$. \\

\noindent 2. Following the $q$-plate, the single photon state becomes
\begin{eqnarray}\label{eq:genpsi}
	1/\sqrt{2}\left(|R\rangle|+1\rangle + |L\rangle|-1\rangle \right),
\end{eqnarray}
as desired.

\section{Tomography:}
Experiments 1 and 2 require that tomography be carried out, once on the full system-environment space of the states $|\psi^\mathrm{NL}_{\cal SE}\rangle$ and $|\psi^\mathrm{L}_{\cal SE}\rangle$, and again on the system alone for each of these states. In both experiments, we consider only a joint SAM-OAM space spanned by the set $\{|R,+1\rangle,|R,-1\rangle,|L,+1\rangle,|L,-1\rangle\}$, which subspace is isomorphous to the two-photon polarization space, and hence, can be represented by the SU(2)$\times$SU(2) group. The Pauli and identity matrices, when combined via tensor product, are generators of the SU(2)$\times$SU(2) group, so that the full SAM-OAM state density matrices $\rho^\mathrm{NL}_{\cal SE}$ and $\rho^\mathrm{L}_{\cal SE}$ can be reconstructed by projecting the birpartite SAM-OAM states of interest onto the eigenvectors of $\hat{\sigma}_x, \hat{\sigma}_y, \hat{\sigma}_z$ and $\hat{\mathbb{I}}$. Therefore, SAM-OAM state reconstruction can be achieved in a manner analogous to the procedure used for two-photon polarization tomography, as long as the OAM states $|+1\rangle$ and $|-1\rangle$ are treated as would be one photon's polarization degree of freedom. Sixteen independent Stokes-like parameters are therefore measured, allowing the unknown state to be unambiguously identified up to the resolution afforded by experimental conditions. In the OAM degree of freedom, these measurements were made by means of a phase-flattening algorithm, which projects the unknown state onto SLMs displaying one of six distinct phase patterns~\cite{nagali:09}. Any phase-flattened light emerging from the SLM's first diffracted order can be coupled to an optical fiber. Tomography carried out on the SAM degree of freedom is effected by using a standard polarization tomography configuration consisting of a quarter- and half-wave plate, followed by a polarizing beam splitter. In concert, these can be used to project an unknown polarization state onto any one of the six polarization states required for SAM tomography.


\begin{thebibliography}{99}
\bibitem{footnote:01}
Foremost among those advancing this more pragmatic perspective is David Mermin, who famously said of the controversy surrounding the Born rule that one should simply, \qo{Shut up and calculate!}
\bibitem{footnote:02}
Here, the notion of perfect correlation is borrowed from Ref~\cite{zurek:05}, in which it is argued that, for a state of the form  $|\psi_{\cal SE}\rangle$, the detection of $|s_{1(2)}\rangle$ implies that the further measurement of any observable on ${\cal E}$ with Schmidt eigenstates will yield a result corresponding to $|\varepsilon_{1(2)}\rangle$, and that, as a result, \qo{partner states} such as $|s_{1(2)}\rangle$ and $|\varepsilon_{1(2)}\rangle$ will be detected together with certainty (\textit{i.e.} with probability $1$).
\bibitem{footnote4}
{To be strictly correct, even the mapping between the physical behaviour of the half-wave plate (and that of the cylindrical lens mode converter) and its mathematical description should be counted as additional assumptions called upon by our investigation.}
\bibitem{footnote:03}
We note that the distinction between entanglement and non-separability is generally emphasized by many who view entanglement as being a nonlocal phenomenon, by definition. We avoid highlighting this distinction, in order to draw a clearer parallel between the mathematical description of the nonlocal entangled state of Experiment 1, and the \qo{local entangled} state of Experiment 2.
\bibitem{footnote:04}
Strictly speaking, envariance can be justified under these conditions with an appeal to the principle of quantum contextuality. Insofar as this principle does not depend upon notions of causality, envariance may be viewed as a generalization of locality and contextuality, whose validity has been demonstrated in its \qo{contextual} and \qo{local} forms by Experiments 1 and 2, respectively.
\end{thebibliography}
\end{document}